\documentclass[sigconf,natbib=true,nonacm]{acmart}
\AtBeginDocument{%
  }

\setcopyright{acmlicensed}
\copyrightyear{2018}
\acmYear{2018}
\acmDOI{XXXXXXX.XXXXXXX}
\acmConference[Conference acronym 'XX]{Make sure to enter the correct
  conference title from your rights confirmation email}{June 03--05,
  2018}{Woodstock, NY}
\acmISBN{978-1-4503-XXXX-X/2018/06}

\begin{document}


\title{Resources for Automated Evaluation of Assistive RAG Systems that Help Readers with News Trustworthiness Assessment}


\author{Dake Zhang}
\orcid{0000-0001-9663-9391}
\affiliation{%
  \institution{University of Waterloo}
  \city{Waterloo}
  \state{Ontario}
  \country{Canada}
}

\author{Mark D. Smucker}
\orcid{0000-0003-4968-6405}
\affiliation{%
  \institution{University of Waterloo}
  \city{Waterloo}
  \state{Ontario}
  \country{Canada}
}

\author{Charles L. A. Clarke}
\orcid{0000-0001-8178-9194}
\affiliation{%
  \institution{University of Waterloo}
  \city{Waterloo}
  \state{Ontario}
  \country{Canada}
}


\begin{abstract}

Many readers today struggle to assess the trustworthiness of online news because reliable reporting coexists with misinformation.
The TREC 2025 DRAGUN (\textbf{D}etection, \textbf{R}etrieval, and \textbf{A}ugmented \textbf{G}eneration for \textbf{U}nderstanding \textbf{N}ews) Track provided a venue for researchers to develop and evaluate assistive RAG systems that support readers' news trustworthiness assessment by producing reader-oriented, well-attributed reports.
As the organizers of the DRAGUN track, we describe the resources that we have newly developed to allow for the reuse of the track's tasks.
The track had two tasks: (Task 1) Question Generation, producing 10 ranked investigative questions; and (Task 2, the main task) Report Generation, producing a 250-word report grounded in the MS MARCO V2.1 Segmented Corpus.
As part of the track's evaluation, we had TREC assessors create importance-weighted rubrics of questions with expected short answers for 30 different news articles.
These rubrics represent the information that assessors believe is important for readers to assess an article's trustworthiness.
The assessors then used their rubrics to manually judge the participating teams' submitted runs.
To make these tasks and their rubrics reusable, we have created an automated process to judge runs not part of the original assessing.
We show that our AutoJudge ranks existing runs well compared to the TREC human-assessed evaluation (Kendall's $\tau = 0.678$ for Task 1 and $\tau = 0.872$ for Task 2).
These resources enable both the evaluation of RAG systems for assistive news trustworthiness assessment and, with the human evaluation as a benchmark, research on improving automated RAG evaluation.

\end{abstract}




\maketitle

\section{Introduction}

Retrieval-Augmented Generation (RAG) has made it practical to build language-model assistants that synthesize information and cite supporting sources~\cite{lewis2020retrieval}.
This shift is especially relevant for tasks where users do not merely want a short answer, but instead need help assembling context and corroborating details across sources.
As RAG systems move from short answers to structured reports, evaluation becomes a central bottleneck.
Human assessment remains the most faithful way to compare systems, yet it is expensive and difficult to reuse when new systems appear.
Meanwhile, standard automatic metrics often fail to reflect whether a generated report actually covers the information a careful reader needs and whether it does so without contradictory claims~\cite{maynez2020on,huang2025survey}.

This paper focuses on the trustworthiness assessment of online news, a setting where the evaluation problem is both socially important and technically subtle.
Online news coexists with misinformation and low-quality reporting, and false or misleading content can spread quickly and widely~\cite{lazer2018science,vosoughi2018spread,allcott2017social}.
Such exposure has been linked to shifts in trust, polarization, and other downstream societal harms~\cite{ognyanova2020misinformation,vasist2023polarizing}.
A growing literature therefore studies interventions and tools that help users slow down, reflect, and better distinguish reliable reporting from deceptive or unsubstantiated content before acting on it~\cite{chan2025cross,guess2020digital,lu2023psychological}.
Yet for many readers, the limiting factor is not access to information but the ability to evaluate it effectively under time constraints, limited domain knowledge, and persuasive presentation tactics~\cite{metzger2013credibility,scharrer2019judging}.

A key insight from research on digital literacy is that expert fact-checkers behave differently from typical readers.
Novices often read \emph{vertically}, staying on the page and relying on on-page cues.
In contrast, professional fact-checkers and skilled readers practice \emph{lateral reading}: they quickly leave the page to investigate the publisher, trace claims to primary sources, and compare coverage across outlets~\cite{wineburg2019lateral,mcgrew2018can}.
This workflow improves accuracy but requires knowing what to check and efficiently finding and weighing evidence.
RAG systems have the potential to assist readers in this news trustworthiness assessment by helping them realize what questions matter for a particular article and providing evidence-grounded context that readers can directly use or inspect further.

Assessing an article's trustworthiness goes beyond checking a set of explicit, well-formed claims.
Most prior automated fact-checking research has framed the problem as a pipeline of claim identification, evidence retrieval, and veracity prediction~\cite{guo2022survey,nakov2021automated}.
Benchmarks such as FEVER~\cite{thorne2018fever} and LIAR~\cite{wang2017liar} have been instrumental in advancing research on evidence retrieval and entailment-style verification.
However, news reporting may be misleading due to selective omission, missing context, or questionable sourcing that are not reducible to a single proposition.
From the reader's perspective, the core task is closer to determining what to investigate and what context they should know to better assess an article's trustworthiness.
This motivates evaluation settings that reward systems for surfacing the most important investigative angles and for producing to-the-point, multi-source, and well-supported reports.

The TREC 2025 DRAGUN (\textbf{D}etection, \textbf{R}etrieval, and \textbf{A}ugmented \textbf{G}eneration for \textbf{U}nderstanding \textbf{N}ews) Track was designed as an end-to-end benchmark for this assistive setting~\cite{zhang2025dragun}.
We defined two complementary tasks over a fixed retrieval corpus (MS MARCO V2.1 Segmented Corpus): (Task 1) generating a ranked list of ten investigative questions that a careful reader should ask of the target news article, and (Task 2, the main task) generating a 250-word report that covers what an informed reader should know to assess its trustworthiness, grounded in retrieved corpus segments with explicit attribution.
The tasks were designed with reader utility in mind: rather than asking systems to output a single \emph{true} or \emph{false} label, DRAGUN requires systems to produce artifacts that support lateral reading by guiding investigation and summarizing corroborating context with verifiable citations.

For evaluation, TREC assessors conducted open-web research and produced importance-weighted rubrics consisting of focused questions with one or more expected short answers, each supported by reference URLs (one example shown in Table~\ref{tab:rubric_example}).
These rubrics specify, for each news article, what an informed reader should know to assess its trustworthiness, and submitted runs were then judged against these rubrics.
This rubric-first design helps the benchmark reflect expert investigative priorities (including angles that submitted systems may miss) and makes scoring interpretable: systems are rewarded for covering high-importance rubric items and penalized when they contradict rubric answers.

DRAGUN builds on the tradition of nugget-based evaluation~\cite{voorhees-2003-evaluating-answers,lin-demner-fushman-2006-will} in that rubric questions and their expected short answers play the role of weighted content units, and the evaluation reduces to judging whether a system's questions and report cover, partially cover, or contradict those units.
To automate nugget-based evaluation, previous work such as Nuggeteer~\cite{marton-radul-2006-nuggeteer} relied on n-gram overlap and corpus-derived term weights to approximate nugget assignments, which was appropriate when robust semantic inference models were not available.
In this paper, we instead leveraged advanced Large Language Models (LLMs) as rubric judges, aligning with recent evidence that LLM-based judges can preserve system rankings and approximate human preferences when appropriately prompted and calibrated~\cite{pradeep2025nuggetizer,assessing2025thakur,zheng2023judging}.
This allows DRAGUN's rubric-based evaluation to scale beyond the original judged submissions.

Accordingly, this resource paper makes DRAGUN reusable by releasing not only the topics, rubrics, and human judgments, but also an automated LLM-based AutoJudge that can score additional runs against the released rubrics.
Our AutoJudge mirrors the human judging protocol using few-shot prompting and produces the same categorical labels used in manual assessment.
When validated against official human judgments, the resulting run-level rankings closely match the human-based rankings (Kendall's $\tau = 0.678$ for Task 1 and $\tau = 0.872$ for Task 2).
Taken together, DRAGUN's released artifacts\footnote{\url{https://github.com/trec-dragun/resources}} enable both (1) benchmarking new assistive RAG systems for news trustworthiness assessment and (2) studying automated evaluation itself, using the human labels and rankings as a reference point:
\begin{enumerate}
    \item A rubric-based benchmark for lateral-reading-style assistance, including 30 news articles and assessor-authored, importance-weighted rubrics of investigative questions with expected short answers and supporting URLs.
    \item Human judgments and scoring code for both question generation and report generation, enabling reproducible evaluation and analysis of system behavior.
    \item An LLM-based AutoJudge that scales rubric-based evaluation to new runs, empirically preserving the official human ranking while keeping the assessments interpretable.
\end{enumerate}

The remainder of this paper describes the DRAGUN tasks (Section~\ref{sec:tasks}), the human rubric construction and assessment procedures (Section~\ref{sec:human-assessment}), the AutoJudge design and validation (Section~\ref{sec:autojudge}), and the released resources and intended reuse scenarios (Sections~\ref{sec:reusable-resources}-\ref{sec:discussion}).

\begin{table}[t]
    \centering
    \caption{Example rubric question with short answers for a news article about the Epic Games vs.\ Apple lawsuit published on \textit{The Verge}.}
    \label{tab:rubric_example}
    \small
    \begin{tabular}{p{0.95\linewidth}}
    \toprule
    \textbf{News Article} \\
    \textbf{Title:} Epic Games is suing Apple \\
    \textbf{URL:} \url{https://www.theverge.com/2020/8/13/21367963/} \\
    \midrule
    \textbf{Question 2:} What is The Verge? \\
    \textbf{Importance:} Have to Know \\
    \midrule
    \textbf{Short Answers:} \\[-4pt]
    \begin{minipage}[t]{0.97\linewidth}
    \begin{enumerate}
        \item The Verge is a technology news website, located in New York City, and owned by Vox Media. \\
        \textbf{\textit{Reference:}} {\footnotesize\url{https://en.wikipedia.org/wiki/The_Verge}}
        \item It is a left-center website with high factual reporting and high credibility. \\
        \textbf{\textit{Reference:}} {\footnotesize\url{https://mediabiasfactcheck.com/the-verge/}}
        \item In the last five years, it has had no failed fact checks. \\
        \textbf{\textit{Reference:}} {\footnotesize\url{https://mediabiasfactcheck.com/the-verge/}}
    \end{enumerate}
    \end{minipage} \\
    \bottomrule
    \end{tabular}
\end{table}

\section{Tasks} \label{sec:tasks}

The TREC 2025 DRAGUN Track consisted of two complementary tasks; participants could complete either or both.
We designed these tasks to work together: Task 1 was to identify critical questions for a given news article, while Task 2 was to create a report that addresses those questions.
This design reflects our core philosophy: we help readers evaluate trustworthiness themselves by providing comprehensive context rather than labeling content as true or false.
We created a website to communicate task descriptions and submission requirements to participants: \url{https://trec-dragun.github.io/}.

\subsection{Corpus and Topics} 

Both tasks used the MS MARCO V2.1 Segmented Corpus\footnote{\url{https://trec-rag.github.io/annoucements/2025-rag25-corpus/\#-ms-marco-v21-segmented-corpus}}, which contains approximately 114 million segments derived from 11 million web documents.
Each segment is a sliding window of 10 sentences with a stride of 5 sentences.
For Task 2, generated reports had to cite relevant segments from this corpus.

We selected 30 news articles from the MS MARCO V2.1 Document Corpus (before segmentation) to serve as \emph{topics}.
These articles were chosen based on two criteria: they cover controversial issues from their publication period (2019-2021, close to the cut-off date when the corpus was collected) or contain content that would attract readers to seek additional context.
The selected articles are from 28 different media sources with diverse political perspectives.
According to Media Bias/Fact Check\footnote{\url{https://mediabiasfactcheck.com/}} ratings in May 2025, our topics include: 13 articles from left-leaning sources, 5 from right-leaning sources, 2 from neutral sources, 4 from pro-science sources, 2 from conspiracy-pseudoscience sources, and 4 from sources without established bias ratings.

\subsection{Task 1: Question Generation} \label{sec:task1-desc}

For each topic (target news article), participants needed to generate 10 critical questions that a thoughtful reader should investigate when assessing the article's trustworthiness, regarding aspects such as source bias, motivation, or alternative viewpoints.
The generated questions were to be ranked from most to least important.
Questions needed to meet the following requirements:

\begin{itemize}
    \item A question's length could not exceed 300 characters.
    \item Questions should not be compound (e.g., \textit{Who is X and when did Y happen?}).
    Each question should focus on a single topic.
    \item Given the context of the article, the questions should not be ambiguous or overly general.
    For example, \textit{Are there other sources corroborating the details presented in this article?} is overly general.
\end{itemize}

\subsection{Task 2: Report Generation}

Report generation is the core task of this track. 
For each news article, participants were to generate a 250-word well-attributed report that provides readers with essential background and context for evaluating trustworthiness.
Report generation can be understood as a RAG task with a fixed query: ``\textit{What should I know about this article to better assess its trustworthiness?}'', but with a varying context, i.e., the news article.
Each sentence of the report could have at most three references (i.e., segment IDs).

\section{Human Assessment} \label{sec:human-assessment}

The track's evaluation of submitted runs was completed by TREC assessors.
The overall process was that during the track participation period, assessors created topic-specific rubrics after conducting open-web research on each article's trustworthiness, using any tools they preferred, e.g., web search.
After runs were submitted, assessors used these rubrics to judge the submitted questions and reports.
We then computed scores based on their judgments.
The remainder of this section provides the details of the rubric construction, question assessment, and report assessment processes.

\subsection{Rubric Construction}

Each topic (news article) was assigned to one primary TREC assessor and two secondary TREC assessors.
Each of the assessors conducted independent research on the trustworthiness of the news article using any tools and resources they deemed appropriate. 
They created rubrics consisting of questions and answers that they thought a good report should cover to help readers determine the trustworthiness of the article.
The primary assessor then merged the rubrics from all three assessors into a single final rubric (capped at ten questions).
The final rubric contains a list of questions, each with one or more short answers.
Table~\ref{tab:rubric_example} shows an example rubric question with expected short answers.
Each question also has an importance label:

\begin{itemize}
    \item \textbf{Have to Know} (4 points): Core, critical questions. 
    Knowing the answer is essential for judging the article's trustworthiness (it might change a reader's perception).
    \item \textbf{Good to Know} (2 points): Important contextual questions.
    Not absolutely critical, but answering them will increase a reader's confidence in their judgment.
    \item \textbf{Nice to Know} (1 point): Background or peripheral questions.
    These provide helpful context but are not crucial for most readers' trust decisions.
\end{itemize}

To standardize rubric quality, we provided assessors with detailed assessment guidelines emphasizing neutral, reader-oriented fact-checking via open-web research and lateral reading.
Concretely, assessors were instructed to investigate (1) the publisher's reputation and potential bias, (2) the author's background, expertise, affiliations, and possible agenda, (3) the veracity of salient claims or statistics made in the article, and (4) the broader context from authoritative reports or research when relevant.
Rubric questions were expected to be focused on a single aspect and phrased without implying an overall verdict on whether the article is true or false.
For each rubric question, assessors wrote one or more concise short answers, and every short answer was required to be backed by at least one supporting reference URL.
We then examined the written rubric answers, and for a few answers that were too long, we shortened them or broke them into multiple short answers.
The guidelines encouraged using credible, English-language, text-based sources that are likely to remain accessible, and not over-relying on a single source. 

\subsection{Question Assessment} \label{sec:human-question-assessment}

The key idea of question assessment is to check how many rubric questions are covered by the participant-submitted questions.
Due to budget constraints, TREC assessors could only judge a subset of the question pairs (rubric question, participant question).
To reduce the pool of question pairs that needed to be judged, we used two models~\cite{zhang2025qwen3embeddingadvancingtext} (\texttt{Qwen3-Embedding-8B}\footnote{\url{https://huggingface.co/Qwen/Qwen3-Embedding-8B}} and \texttt{Qwen3-Reranker-8B}\footnote{\url{https://huggingface.co/Qwen/Qwen3-Reranker-8B}}), each of which independently selected the most similar/relevant participant question for each rubric question.

For each rubric question, the assigned primary assessor judged one (if both models picked the same question) or two questions from the participant's question list (ten questions), and assigned one of the following four similarity labels to each question pair:

\begin{itemize}
    \item \textbf{Very Similar}: Questions may have different wording, but answering either question provides effectively the same information to the reader.
    \item \textbf{Similar}: Answering the questions will provide similar, but slightly different information to the reader.
    \item \textbf{Different}: The answer to each question will provide different information, with possibly some overlap, to the reader.
    \item \textbf{Very Different}: Answers to questions provide different information, with little to no overlap, to the reader.
\end{itemize}

We scored runs using these assessor-assigned labels ($\ell$).
For topic $t$, let $\mathcal{Q}_t$ be rubric questions, and let
$w_q \in \{4,2,1\}$ be the importance weight of rubric question $q\in\mathcal{Q}_t$.
For run $r$, let $\mathcal{P}_{r,t}$ denote the set of submitted questions.
We mapped assessor labels to numeric similarity scores as:
\begin{equation*}
g(\ell) =
\begin{cases}
1 & \ell = \textsc{very similar},\\
0.5 & \ell = \textsc{similar},\\
0 & \ell \in \{\textsc{different}, \textsc{very different}\}.
\end{cases}
\label{eq:t1_label_map}
\end{equation*}

Let $\ell_{r,t}(q,p)$ be the judged label for rubric question $q$ and submitted question $p$.
We score run $r$ on topic $t$ by rewarding only the best-matching submitted question per rubric question:
\begin{equation*}
S_{r,t}
= \frac{1}{W_t}
\sum_{q \in \mathcal{Q}_t} w_q \, \max_{p \in \mathcal{P}_{r,t}} g\big(\ell_{r,t}(q,p)\big),
\ \ \ \
W_t = \sum_{q \in \mathcal{Q}_t} w_q.
\label{eq:t1_score}
\end{equation*}

Unjudged pairs were assigned \textsc{very different} (zero credit).
In this setting, the same submitted question could be labeled as similar to multiple rubric questions and rewarded by design, as we wanted to measure the rubric coverage of participant questions.

Regarding the second requirement mentioned in Section~\ref{sec:task1-desc}, i.e., the exclusion of compound questions, this constraint was not explicitly enforced during the human assessment phase.
To ensure adherence to task guidelines, we used \texttt{gpt-oss-120b} to automatically identify and filter out compound questions (11.3\%). 
We validated this classification approach by manually labeling a stratified sample of 100 compound and 100 non-compound questions identified by the model. 
The classifier demonstrated high reliability, achieving a True Positive Rate of 0.989 and a False Positive Rate of 0.124.
With this satisfactory performance, we adopted this automated filtering mechanism to remove compound questions from the submitted question list ($\mathcal{P}_{r,t}$) before scoring.
We acknowledge a limitation in this post-hoc filtering approach.
Because the selection of participant questions for human judgment occurred prior to the removal of compound questions, it is possible that a compound question was selected as the most similar match for a rubric question.
If such a question were subsequently filtered out, the run would receive no credit for that specific rubric question, even if a valid, non-compound alternative existed with lower similarity scores.

\subsection{Report Assessment} \label{sec:human-report-assessment}

Similar to question assessment, the key idea of report assessment is to check how many rubric answers are covered by each report.
These rubric answers function as exemplars of the key information we want a report to convey, rather than templates that systems must match verbatim.
For topic $t$, each rubric question $q\in\mathcal{Q}_t$ has a set of short answers $\mathcal{A}_{t,q}$.
For a report from run $r$, the primary assessor assigned a label $\ell_{r,t}(a)\in\{\textsc{supports},\textsc{partial},\textsc{contradicts},\textsc{none}\}$
to each rubric answer $a$, i.e., using rubric answers as a checklist.

\begin{itemize}
    \item \textbf{Supports}: The report provides an answer to the question, consistent with the key elements of the rubric answer.
    \item \textbf{Partial}: The report provides an answer to the question that contains some but not all key elements of the rubric answer.
    \item \textbf{Contradicts}: The report contains information that contradicts the rubric answer. 
    If the report supports or has partial support for the rubric answer, but it also contradicts the answer, then the \emph{contradicts} label takes precedence.
    \item \textbf{None}: The report has no support or connection with the rubric answer. This is the default.
\end{itemize}

We defined label-to-value mappings for \emph{supportive} and \emph{contradictory} scoring:
\begin{equation*}
v_{\mathrm{sup}}(\ell) =
\begin{cases}
1 & \ell=\textsc{supp.},\\
0.5 & \ell=\textsc{part.},\\
0 & \ell\in\{\textsc{contr.}, \\
  & \quad \quad \textsc{none}\},
\end{cases}
\ \ \ \
v_{\mathrm{con}}(\ell) =
\begin{cases}
1 & \ell=\textsc{contr.},\\
0 & \ell\in\{\textsc{supp.},\textsc{part.},\\
  & \quad \quad \textsc{none}\}.
\end{cases}
\label{eq:t2_label_map}
\end{equation*}
We computed topic-level supportive and contradictory scores as:
\begin{align*}
S^{\mathrm{sup}}_{r,t}
&= \frac{1}{W_t}
\sum_{q\in\mathcal{Q}_t} \frac{w_q}{|\mathcal{A}_{t,q}|}
\sum_{a\in\mathcal{A}_{t,q}} v_{\mathrm{sup}}\!\big(\ell_{r,t}(a)\big),
\\
S^{\mathrm{con}}_{r,t}
&= \frac{1}{W_t}
\sum_{q\in\mathcal{Q}_t} \frac{w_q}{|\mathcal{A}_{t,q}|}
\sum_{a\in\mathcal{A}_{t,q}} v_{\mathrm{con}}\!\big(\ell_{r,t}(a)\big),
\end{align*}
where $W_t=\sum_{q\in\mathcal{Q}_t} w_q$.
Ranking of runs is by the average supportive score over 30 topics (higher is better); contradictory (lower is better) is reported as a secondary diagnostic.

\subsection{Per-topic Scores and Headroom}

\begin{figure}
  \centering
  \includegraphics[width=0.98\linewidth]{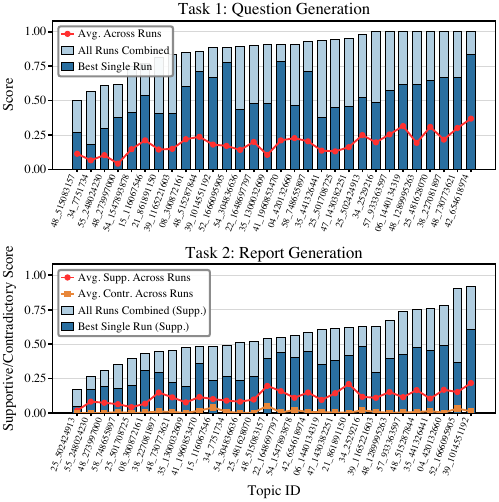}
  \caption{Per-topic rubric-weighted scores. Dark bars: best run for each topic; light: pooled upper bound. Topic IDs omit the \texttt{msmarco\_v2.1\_doc\_} prefix.}
  \Description{Two stacked bar charts showing per-topic scores for the DRAGUN track. The top panel (Task 1: Question Generation) and bottom panel (Task 2: Report Generation) each display a light blue bar for the all-runs-combined upper bound and an overlaid dark blue bar for the best single run, sorted by the upper bound. A red line shows the average score across all runs per topic. The bottom panel additionally includes an orange line for the average contradictory score, which remains near zero across all topics.}
  \label{fig:per-topic-scores}
\end{figure}

Figure~\ref{fig:per-topic-scores} summarizes per-topic rubric-weighted scores for the runs submitted to the TREC 2025 DRAGUN track.
For each topic, the dark bar shows the best run, and the light extension shows additional headroom up to a ceiling formed by pooling, for each rubric element, the strongest match across all submissions (``All Runs Combined'').
The lines indicate mean scores across runs; for Task~2, we plot both the mean supportive score (red) and the mean contradictory score (orange).

For Task~1 (Question Generation), scores vary substantially across topics, and the best-performing run differs by topic.
At the same time, the pooled ceiling is high for most topics and, because each rubric is capped at ten questions, the corresponding ``super run'' still satisfies the 10-question submission constraint.
This suggests that the rubrics largely reflect investigative directions that current systems can generate, but that different systems cover different subsets of these directions.
The gap between the best single run and the pooled ceiling, therefore, represents feasible headroom for more robust, article-adaptive question generation.

For Task~2 (Report Generation), supportive coverage is consistently lower than in Task~1, reflecting the added difficulty of retrieving evidence and compressing it into a 250-word, well-attributed report.
The gap between the best run and the pooled ceiling indicates clear headroom in retrieval, evidence selection, and writing under a strict length budget.
Unlike Task~1, this pooled ceiling is generally not attainable by a single report, since combining all supported elements would typically exceed the length limit, so it should be interpreted as an information-availability ceiling across existing systems.
Importantly, the average contradictory scores are much smaller than the supportive scores (orange vs.\ red), suggesting that explicit contradictions to rubric answers are not a major issue for existing runs and do not meaningfully distinguish them.
Nevertheless, contradiction labels remain informative: they flag report content that is undesirable to include, even if current differences between runs are driven primarily by what rubric-relevant information systems support rather than what they contradict.

\section{AutoJudge: Reusable Automatic Assessment} \label{sec:autojudge}

\begin{figure}
  \centering
  \includegraphics[width=\linewidth]{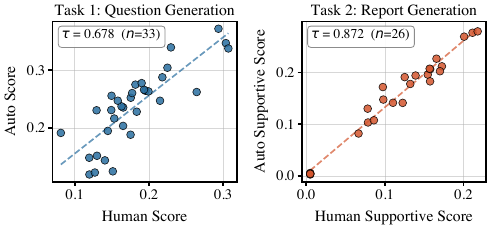}
  \caption{Correlation between human and LLM-based automatic assessments at the run level (averaged score across 30 topics). Each point represents a run.}
  \Description{Two scatter plots displayed side by side. The left plot shows human scores versus automatic scores for 33 question-generation runs, with points clustered along an upward-sloping dashed trend line and a Kendall's tau of 0.678. The right plot shows human supportive scores versus automatic supportive scores for 26 report-generation runs, with points following a steeper upward trend and a Kendall's tau of 0.872. Both plots demonstrate a strong positive correlation between human and automatic rankings.}
  \label{fig:human-vs-auto}
\end{figure}

The DRAGUN rubrics and human judgments provide a high-quality but \emph{static} evaluation of the 2025 submitted runs.
To make the collection reusable for evaluating \emph{future} systems on the same tasks, we additionally provide an AutoJudge that can score new runs against the rubrics without requiring new assessor effort.
Recent work in IR has shown that LLMs can serve as effective judges for rubric-like assessment and can preserve system rankings derived from humans (e.g., by reporting high rank correlation using Kendall's $\tau$) \cite{assessing2025thakur, pradeep2025nuggetizer}.

We implemented a few-shot AutoJudge using OpenAI's open-weight \texttt{gpt-oss-120b}\footnote{\url{https://openai.com/index/introducing-gpt-oss/}} with \texttt{temperature} set to 0 and \texttt{top\_p} set to 1 (other capable LLMs should also work).
The prompt mirrors the assessment instructions used for TREC assessors and includes labeled examples from our organizer baseline runs~\cite{zhang2025baseline} (4 baselines for Task~1 and 2 baselines for Task~2, excluded from later validation).
The model outputs the same labels as the human judging protocol, and we computed run scores using the same scoring scripts as Sections~\ref{sec:human-question-assessment} and~\ref{sec:human-report-assessment}.
On an NVIDIA RTX PRO 6000 GPU, it took roughly 13 hours to assess all 77,880 question pairs and 780 reports.
For validation, we report (1) run-level rank correlation via Kendall's $\tau$ and (2) label-level agreement between AutoJudge and the human labels using Cohen's $\kappa$ \cite{cohen1960kappa} and Gwet's AC1 \cite{gwet2008ac1}.
We report AC1 because $\kappa$ is known to be deflated under heavy class imbalance (the prevalence paradox)~\cite{FEINSTEIN1990543, BYRT1993423}.

For each rubric question in Task 1 evaluation, we showed the judge the target article, the rubric question, and 40 baseline questions with assessor-provided similarity labels: \emph{very similar}, \emph{similar}, \emph{different}, or \emph{very different}.
Baseline questions not selected for human judging were treated as \emph{very different}, matching the track scoring where non-judged pairs received zero credit.
The AutoJudge then labeled each (rubric question, participant question) pair, and we computed rubric-weighted run scores after filtering out compound questions.
At the run level, the automatic ranking moderately matches the official human ranking (Kendall's $\tau=0.678$, $n=33$; Figure~2 left).
At the label level, we collapsed \emph{different} and \emph{very different} into a single \emph{no-credit} class (since both yield 0 points) and obtained 82.1\% raw agreement, with $\kappa=0.472$ and AC1 $=0.785$.

In terms of report evaluation, for each topic, we provided the judge with the target article, the complete rubric (questions and short answers), and two baseline reports labeled at the rubric-answer level.
For a candidate report, the judge assigned one of \{\emph{supports, partial, contradicts, none}\} for each rubric answer, and we computed the supportive score using the same procedure as Section~\ref{sec:human-report-assessment} (contradictory scores are similarly low across runs, and therefore not distinguishable).
The resulting run-level ranking is highly aligned with the human-derived ranking (Kendall's $\tau=0.872$, $n=26$; Figure~2 right).
At the answer level, we observed 86.7\% raw agreement, with $\kappa=0.50$ and AC1 $=0.85$.

Our rank correlations are in line with recent IR auto-judging studies that compare LLM-derived and human-derived system orderings.
For example, LLM-based support judging in the TREC 2024 RAG Track reports run-level Kendall's $\tau$ around 0.8 \cite{assessing2025thakur}, and AutoNuggetizer-style nugget evaluation reports run-level Kendall’s $\tau$ between 0.727 and 0.901 depending on the manual reference condition \cite{pradeep2025nuggetizer}.
Overall, our AutoJudge provides the missing component needed to make the released rubrics a reusable benchmark: future systems can be scored consistently with the official human-evaluated leaderboard, enabling direct follow-up experimentation beyond the original track runs.

\section{Reusable Resources} \label{sec:reusable-resources}

To support replication of the DRAGUN track and follow-up research on assistive RAG for news trustworthiness assessment, we release a reusable package of data, judgments, and evaluation code.
Most artifacts are available in our public GitHub repository (\url{https://github.com/trec-dragun/resources}).
The repository README documents the terms of use, citation instructions, file formats, and directory structure for all released artifacts, along with detailed tutorials on how to use the released package.
For resources that are conventionally hosted elsewhere (e.g., raw run files on TREC's website\footnote{\url{https://pages.nist.gov/trec-browser/}}), the repository provides pointers and detailed instructions for obtaining them.
Table~\ref{tab:statistics} summarizes the collection size and composition.
The released package includes the following components:
\begin{itemize}
    \item Topic set and assessor rubrics.
    \begin{itemize}
        \item Topics (30 news articles): the set of target articles selected from the MS MARCO V2.1 Document Corpus.
        \item Rubrics (30): importance-weighted rubrics created by TREC assessors, one per topic, consisting of questions with one or more expected short answers. Each short answer is supported by one or more reference URLs.
    \end{itemize}
    \item Human judgments, with assessment guidelines.
    \begin{itemize}
        \item Task 1 (Question Generation): assessor judgments of question similarity between rubric questions and participant questions.
        \item Task 2 (Report Generation): assessor judgments of whether a report supports, partially supports, contradicts, or does not address each rubric answer.
    \end{itemize}
    \item Participant submissions, with participation guidelines.
    \begin{itemize}
        \item Runs: all participating teams' submissions.
        \item Baseline system: an iterative multi-agent baseline RAG system implementation covering both tasks.
    \end{itemize}
    \item LLM-based AutoJudge.
    \begin{itemize}
        \item AutoJudge system: a few-shot prompting-based LLM judge (using \texttt{gpt-oss-120b}) that can score additional runs beyond those assessed during the original track.
        \item LLM-based assessments: Assessments from the AutoJudge of existing runs for both tasks.
    \end{itemize}
    \item Scoring scripts: Python scripts that compute scores from either the human judgments or the LLM-based assessments.
\end{itemize}

\begin{table}
    \centering
    \caption{Statistics of the DRAGUN collection.}
    \label{tab:statistics}
    \footnotesize
    \setlength{\tabcolsep}{4pt}
    \begin{tabular}{lr}
    \toprule
    Topics (News Articles) with Rubrics & 30 \\
    Rubric Questions (Avg / Rubric: 7.9) & 236\\
    Rubric Answers (Avg / Rubric: 18.4; Avg / Question: 2.3) & 551 \\
    \midrule
    \textbf{Task 1: Question Generation} & \\
    Submitted Runs / Teams / Questions & 37 / 10 / 11,100 \\
    \ \ \ \ Compound Questions / Not Compound Questions& 11.3 / 88.7 (\%) \\
    Total Rubric-participant Question Pairs & 87,320 \\
    Human-assessed Question Pairs & 12,733 \\
    \ \ \ \ Very Similar / Similar / Different / Very Different 
      & 8.7 / 12.9 / 16.9 / 61.5 (\%) \\
    \midrule
    \textbf{Task 2: Report Generation} & \\
    Submitted Runs / Teams / Reports & 28 / 8 / 840 \\
    Answer-report Pairs (All Human-assessed) & 15,428 \\
    \ \ \ \ Supports / Partial / Contradicts / None 
      & 5.0 / 7.8 / 0.8 / 86.4 (\%) \\
    \bottomrule
    \end{tabular}
\end{table}

\section{Discussion} \label{sec:discussion}

Our release supports several follow-up research directions.

\textbf{Benchmarking future systems.}
The most direct use is to evaluate new systems without requiring additional assessor effort.
A key design decision was to have assessors build rubrics using open-web lateral reading, rather than restricting rubric construction to the MS MARCO V2.1 Segmented Corpus.
This choice helps ensure that rubrics capture comprehensive and unbiased information beyond what submitted systems retrieve. 
It avoids a common pitfall of ``pool-nuggets-then-judge'' RAG evaluation, where the evaluation target (nuggets) is constrained by what participating systems happen to surface and fail to spot.
The tradeoff is that some rubric answers may be absent from a fixed retrieval corpus, but the headroom analysis suggests substantial missing coverage even for information that should be discoverable (Figure~\ref{fig:per-topic-scores}), making these rubrics useful for diagnosing retrieval and synthesis gaps.

\textbf{Comparing evaluation norms.}
Our rubric-first workflow (expert rubric creation followed by rubric-based scoring) enables direct comparison with nugget-oriented paradigms that derive evaluation units from pooled system outputs, such as AutoNuggetizer~\cite{pradeep2025nuggetizer} and RUBRIC~\cite{farzi2024pencils}.
DRAGUN makes it possible to quantify how rankings change under these different norms and to study whether report-derived nuggets systematically miss expert-identified angles that matter for news trustworthiness assessment.

\textbf{Advancing automated judging for RAG systems.}
Because DRAGUN provides both expert labels and an LLM-based AutoJudge, it can serve as a benchmark for developing stronger judges that better match expert decisions, both at the label level and in preserving system rankings.
Our current report evaluation is only on rubric-answer coverage (support and contradiction).
Future work could extend the framework with complementary dimensions that are important for RAG, such as citation faithfulness (whether cited evidence actually supports the associated claim), readability, etc.

\textbf{Scaling rubric creation and using rubrics for training.}
Finally, DRAGUN demonstrates that expert-authored rubrics are a feasible and effective means of specifying the critical information in a news trustworthiness report.
By successfully implementing a complete pipeline, from drafting assessment guidelines to training TREC assessors in lateral-reading techniques, we have obtained high-quality rubrics capable of providing nuanced evaluation for RAG systems.
This pipeline can be scaled up to generate larger benchmarks and diverse training datasets, providing the necessary supervision signal to align LLM-based assistants with expert-level investigative behaviors.
Such scaling would enable models to retrieve, cite, and synthesize evidence addressing expert-identified angles while penalizing contradictory or unsubstantiated claims.

\section{Conclusion} \label{sec:conclusion}

We have made the TREC 2025 DRAGUN Track into a reusable resource for the evaluation of assistive RAG systems that help readers assess news trustworthiness.
Our release packages news articles with importance-weighted rubrics, participant submissions, and human judgments for both Task 1 (Question Generation) and Task 2 (Report Generation).
To support evaluation beyond the originally judged runs, we have also created and provided an LLM-based AutoJudge that mirrors the rubric-based judging protocol and produces rubric-coverage labels at the same granularity as the human assessments.
When validated against the official human judgments, AutoJudge preserves the run-level ordering well, achieving Kendall's $\tau = 0.678$ for Task~1 and $\tau = 0.872$ for Task~2.
Together, these resources enable reproducible benchmarking of future systems for lateral-reading-style assistance, and they provide a concrete testbed for advancing automated RAG evaluation using expert rubrics and human labels as a reference point.

\begin{acks}

This research received funding from Microsoft and the Natural Sciences and Engineering Research Council of Canada (NSERC) grant ALLRP/597573-24. 
We extend our appreciation to all participants who submitted work to the 2025 track.
We are particularly grateful to Ian M.\ Soboroff and Hoa T.\ Dang at NIST for their coordination of the run assessment process, and to all TREC assessors who contributed their expertise to the evaluation.

\end{acks}

\bibliographystyle{ACM-Reference-Format}
\bibliography{references}




\end{document}